\documentclass[runningheads]{llncs}
\usepackage[T1]{fontenc}
\usepackage{xcolor,pifont}
\newcommand*\colourcheck[1]{%
  \expandafter\newcommand\csname #1check\endcsname{\textcolor{#1}{\ding{52}}}%
}

\colourcheck{blue}
\colourcheck{green}
\colourcheck{red}

\newcommand*\colourhatch[1]{%
  \expandafter\newcommand\csname #1hatch\endcsname{\textcolor{#1}{\ding{55}}}%
}

\colourhatch{blue}
\colourhatch{green}
\colourhatch{red}

\usepackage{graphicx}
\usepackage{longtable}
\usepackage{multirow}
\usepackage{rotating}
\usepackage{longtable,multirow,graphicx}

\usepackage{url}
\begin{document}
\title{Assessing Spear-Phishing Website Generation \\ in Large Language Model Coding Agents}
\titlerunning{LLM Spear-Phishing}
%
%
\author{Tailia Malloy\and Bissyande F. Tegawende}
\authorrunning{Malloy and Bissyande}
\institute{University of Luxembourg, Interdisciplinary Center for Security, Reliability, and Trust (SnT), Trustworthy Software Engineering Group (TruX), Luxembourg}

\maketitle

\begin{abstract}
    Large Language Models are expanding beyond being a tool humans use and into independent agents that can observe an environment, reason about solutions to problems, make changes that impact those environments, and understand how their actions impacted their environment. One of the most common applications of these LLM Agents is in computer programming, where agents can successfully work alongside humans to generate code while controlling programming environments or networking systems. However, with the increasing ability and complexity of these agents comes dangers about the potential for their misuse. A concerning application of LLM agents is in the domain cybersecurity, where they have the potential to greatly expand the threat imposed by attacks such as social engineering. This is due to the fact that LLM Agents can work autonomously and perform many tasks that would normally require time and effort from skilled human programmers. While this threat is concerning, little attention has been given to assessments of the capabilities of LLM coding agents in generating code for social engineering attacks. In this work we compare different LLMs in their ability and willingness to produce potentially dangerous code bases that could be misused by cyberattackers. The result is a dataset of 200 website code bases and logs from 40 different LLM coding agents. Analysis of models shows which metrics of LLMs are more and less correlated with performance in generating spear-phishing sites. Our analysis and the dataset we present will be of interest to researchers and practitioners concerned in defending against the potential misuse of LLMs in spear-phishing.
\end{abstract}

\section{Introduction}
Spear-phishing involves personally tailored methods of social engineering that target a specific subpopulation such as the members of a single industry, company, or even an individual. Traditionally, spear-phishing has been highly labor intensive, requiring research into individuals or companies and the creation of social engineering attempts that are more likely to deceive those groups \cite{burns2019spear}. Generative Artificial Intelligence (GAI) models such as Large Language Models (LLMs) have the potential to significantly reduce the time and effort requirements of designing spear-phishing attempts \cite{bethany2024large} including generating websites tailored to specific companies or individuals \cite{toth2024llms}. This is enabled by extensions that integrate LLMs into Integrated Development Environments (IDEs) which allows them to control more complex programming environments like websites, networks, and databases \cite{tan2023copilot}. The programming abilities of these agents combined with their web-searching capabilities allows them to gather real-time up to date information on individuals and companies that have publicly available information on the internet \cite{nam2024using}. Taken together, these features of LLM coding agents raise the concern over their impact on cybersecurity, and we specifically are interested in the generation of spear-phishing websites. 

This work exists in the context of previous studies that have compared the ability of LLMs in social-engineering contexts with human subjects. One previous experiment has demonstrated the ability of LLMs to generate HTML and CSS code that can effectively deceive humans \cite{malloy2025training} more effectively than emails that are generated by human work alone. Further research has described methods of improving anti-phishing training using LLMs to generate training examples and educational feedback in an online educational platform. Part of the reason for this improvement in educational quality by leveraging LLMs themselves is due to the ability of LLMs to tailor educational feedback to individuals, leading to higher educational outcomes \cite{malloy2025you}. However, agentic AI coding agents have the potential to tailor social-engineering attempts to individuals, raising the question of whether this type of education remains effective when LLMs are generating user specific social engineering attempts. 

In this work, we assess the capability of LLMs in their ability to generate individual specific spear phishing emails and websites. The purpose of this is to raise awareness about the issues surrounding the misuse of LLMs. Additionally, we are interested in improving the general knowledge around methods for safety in LLM coding agents, to prevent their misuse. To that end, we release a dataset of LLM generated websites that can be used in the future for a variety of methods of improving online security. Additionally, we compare various methods for enabling the generation of these websites by LLMs that are designed with safety features intended to reduce the likelihood of generating this type of output. Again, this is done to inform the research community about the limitations of LLM safety. To prevent the misuse of these methods of overcoming LLM safety features in the design of spear-phishing campaigns, we also do not detail in full the methods used. This project highlights the potential misuse of LLMs by cyberattackers in the targeting of specific individuals through spear-phishing campaigns. These results will be useful to researchers interested in improving the quality of spear-phishing detection training. Code and data for this work is available at\footnote{\url{https://github.com/TailiaReganMalloy/assessingSpearPhishing}}

\subsection{Spear-Phishing}
Spear-Phishing (SP) is a method of gaining access to sensitive and private information or installing malware onto target computers by tricking specific individuals into clicking links, sending credentials, or otherwise giving access to unauthorized users. The specificity of SP differentiates itself from other types of Social Engineering like Phishing, which is done using generic targets and sent to a wide variety of targets. Examples of SP with messaging include writing emails, texts, or direct messages in social media applications. These messages are designed to be visually similar to the messages a specific target group may receive. Other instances of SP include creating entire websites that replicate a website that is used by a company or individual, which can allow for the collection of passwords and other sensitive information. While SP can be more effective than traditional generic approaches to Phishing \cite{butavicius2016breaching}, it is more costly and time consuming to develop due to the requirements of making specialized campaigns that may only target a small group such as an individual CEO \cite{burns2019spear}. 

One of the more time-intensive methods for SP is the creation of websites that are used to replicate real sites that are used by companies or a small number of individuals \cite{mohammad2015phishing}. These can be used alongside emails or other messages that are written for a single individual, referred to as highly targeted spear phishing attacks \cite{burns2019spear}. These targeted SPs can be designed using publicly available information to convince users to trust the source of the phishing email or the phishing website. While this approach can be costly in both time and money, it has been demonstrated to be highly effective and difficult to train users to recognize \cite{ayoola2024effectiveness}. However, the significant costs of writing code to replicate real websites and hosting these sites has limited their broad applicability. One of the major concerns of LLM Agents that can generate code is the increase in potential applicability of SP, which is discussed in the next section.

\subsection{LLM Agent Code Generation}
LLM Agents are a use case of LLMs and AI systems that rely on LLMs to extend their applicability beyond being a tool for users into an agent that can take actions in an environment, impact that environment, and observe and reason about the changes in the environment resulting from their actions. This can be done by instructing LLMs to reply with text in a specific format such as JSON that can be parsed and fed into software to trigger specific commands  \cite{tang2024worldcoder}. These commands can include creating files, downloading software packages, or changing environment variables \cite{wang2024executable}. This is done in an attempt to make LLM Agents that are autonomous and require little to no human intervention in their application onto a variety of tasks \cite{fang2024llm}. Instead of only outputting text and answering questions, LLM agents output commands that can be run by software to perform a variety of actions depending on the task and environment. 

In the context of automated code generation, LLMs can create and edit files, run-time environments, environment variables, programming packages, and more. According to the 2024 Q3 Earnings call made by Alphabet Inc, "Today, more than a quarter of all new code at Google is generated by AI, then reviewed and accepted by engineers." \cite{GoogleEarnings}. LLMs are also used in tandem with work from human programmers such as by checking for bugs in code \cite{tang2024codeagent} and their potential solutions. While these methods are very effective, they have the risk of being misused by cyberattackers to design and disseminate Social Engineering attacks like SP. One of the biggest concerns for the misuse of LLMs in this way is that it greatly expands the number of people who are able to launch SP attacks, while also reducing the time, effort, and cost associated with creating these code bases \cite{tang2025autoagent}. The following section will detail how LLM Agents can and are being used in this way. 

\subsection{LLM Social Engineering}
As LLMs have become more and more human-like in the text they are able to generate, the ability for people to distinguish between when they are talking to a real human or a chatbot has deteriorated \cite{jones2025large}. This has important implications for the field of social engineering research which seeks to understand methods used by attackers to leverage human social skills to convince users to reveal their personal information or give access to their systems \cite{kumarage2025personalized}. LLMs are able to imitate common methods for social engineering such as the use of a sense of authority to convince users that they need to follow the direction of the attacker, or sending offers that entice users to reply in order to receive payment or other goods \cite{kumarage2025personalized}. LLM social engineering exists in the context of a long history of automated social engineering that has sought to make it easier for attackers to send a wide net of social engineering attempts through mass emails or messages, to have the best chance of reaching a vulnerable target \cite{huber2009towards}. However, what makes LLM social engineering attacks unique among these approaches is the ease of use for non-experts that is enabled by simple to use and publicly available LLM systems. 

While LLMs pose a significant threat for misuse, they can also be leveraged in the defense against social engineering. One approach to this is by leveraging LLMs to classify and detect phishing attempts and other forms of social engineering in an effort to prevent these messages from being sent to users \cite{roumeliotis2024next}. Alternatively, LLMs can be used to highlight or give context to potentially dangerous emails so that users are aware of the contents of the messages they receive. Education using LLMs can also be significantly beneficial due to the low cost of LLM based educational platforms compared to training using in-person instruction \cite{malloy2025training}. These educational platforms can be more interactive than books or static web-pages, which can aid in the retention of information while still being low cost and widely available. Recent methods have applied LLMs to train students to identify these types of generated emails by generating example phishing emails and enabling a chat-bot to give feedback during training \cite{malloy2025improving}.

\subsection{Spear-Phishing with LLMs}
Recent research has highlighted the potential for LLMs to be misused for spear-phishing. One study explored this by using GPT-3.5 and GPT-4 to generate spear-phishing messages for 600 members of the British Parliament using prompt engineering to bypass the safety mechanisms built into these models \cite{hazell2023large}. This work focused on the reconnaissance phase, where cyberattackers narrow down who they will send spear-phishing attempts to from a large pool of potential victims, as well as the message generation phase, where personalized messages are generated based on information that is available about them. Generating these personalized messages can be highly cost efficient compared to traditional spear-phishing methods \cite{hazell2023spear}, but this raises the question of how effective LLM generated spear-phishing attempts are. A study of 9000 individuals at a large university found that LLM generated Lateral (originating from the same organization) spear-phishing emails were as effective as ones created by communications professionals \cite{bethany2024large}. Additional work has focused on the applicability of LLM agents specifically, finding that existing automated systems for spear-phishing detection in email applications as well as LLM-based detection strategies were insufficient for defense against attacks that were rephrased by LLM agents \cite{afane2024next}

As with other social engineering methods, LLMs have also been employed in the detection and defense against spear-phishing attempts. An approach to spear-phishing detection relying on the embeddings formed by an ensemble of LLMs to categorize messages resulting in 91\% accuracy in classification \cite{nahmias2024prompted}. Methods for detecting LLM generated spear-phishing have also introduced datasets of emails, such as the \textit{SpearMail} dataset which contains 14,672 LLM generated spear-phishing emails based on 681 public profiles, which can be used for benchmarking and comparing detection techniques \cite{liu2025pimref}. The dataset presented in this work differs from this and related previously created datasets in that it contains repositories of JavaScript, HTML and CSS code used for websites that are intended to be used in spear-phishing campaigns. Much of the previous research on LLM spear-phishing has focused only on the emails and other messages that are sent to targets, without focusing on the area of spear-phishing websites \cite{weinz2025impact,opara2025evaluating,bethany2024large,afane2024next,heiding2024devising}. This is an important aspect of spear-phishing, as having a target simply click on a link in an email is often not enough to compromise their system due to advanced security techniques that verify the legitimacy of links and prevent automatic downloads or other vectors of attack.

\section{Methods}
\subsection{Assessment Pipeline}
\begin{figure}[ht]
  \centering
  \includegraphics[width=0.9\textwidth]{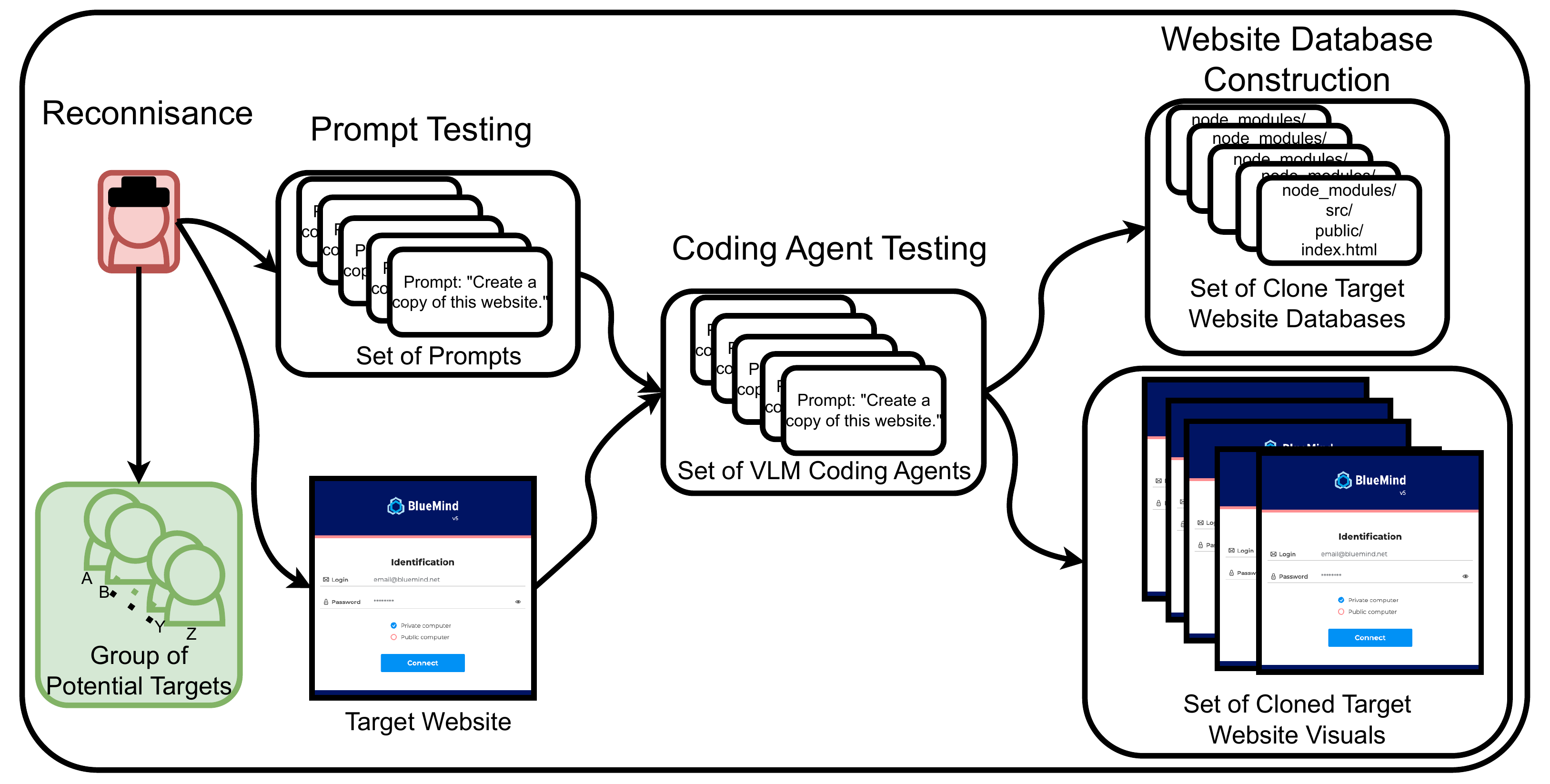}
  \caption{A pipeline for assessing LLM generated spear-phishing websites and messages.} 
  \label{fig:AssessmentPipeline}
\end{figure}

We are interested in assessing different LLM coding agents for their ability and willingness to generate spear-phishing websites primarily in no-code or low-code settings, where the end-user is assumed to have low to no technical programming skills and relies on natural language in their prompting \cite{wang2023natural}. This setting is of interest to us because of the major concern of the potential of LLM coding agent misuse by cyberattackers greatly expanding the ease of spear-phishing campaign generation. This is an important first step in the defense against LLM generated spear-phishing, as it can allow for the development of automated systems for detecting these types of websites, as well as educational materials and platforms that are useful for aiding end-users in understanding the dangers posed by LLMs. To that end, we developed the LLM spear-phishing pipeline shown in Figure \ref{fig:AssessmentPipeline}, which details the steps that a cyberattacker may use in reconnaissance, website development, message generation, and spear-phishing campaign launching. We assess LLM coding agents in their willingness and effectivity in each of these areas, but focus specifically on the website development as it is the least well studied of these four areas. 

The first phase of our assessment pipeline is a cyberattacker doing reconnaissance to determine the website that will be recreated using an LLM coding agent. This is an often ignored type of reconnaissance, which typically focuses on the later step of selecting a subset of targets to send spear-phishing messages to. But since this work focuses on website generation, it is an important part of the type of spear-phishing we are interested in. LLM agents can be used in this phase primarily to determine which websites are used by the target group, and which ones can most easily be recreated by the LLM coding agent as a part of a successful spear-phishing campaign. After determining what website will be recreated, the pipeline continues onto the creation of multiple different prompts to perform prompt testing that will be used with the set of LLM coding agents. In our case, we use Visual Language Models (VLMs) that can take in as input both text and visual information. This allows for the prompting to include both the text request to design the website and the target website image as reference, to make the generated website as similar as possible. After prompting all models with the set of prompts and target website, the pipeline continues onto the construction of a database of website clones, which can be used to generate images of the cloned target website that will be used to evaluate which code base will be used in the spear-phishing campaign. 

\subsection{Code Generation}
All code in our dataset is generated using the Microsoft Visual Studio Code integrated development environment with the GitHub Copilot Chat extension installed which allows for LLMs to be integrated into the development environment through a series of commands that are available to it that can be used to create files, run scripts, view script output and errors in the terminal, among other functions\footnote{\url{https://code.visualstudio.com/docs/copilot/overview}}. All GPT, Claude, and Gemini models are directly accessible by the Copilot extension and were used directly through it, with all remaining LLMs being accessed through the OpenRouter platform which allows for routing of LLM requests through the relevant application programming interface for the remaining models\footnote{\url{https://openrouter.ai/docs/quickstart}}. While this Copilot chat extension allows for LLMs to output commands that can control the programming environment, models that are not specifically trained or fine-tuned may have a more challenging time doing this efficiently. Indeed, a review of the databases and chats in out presented dataset shows that some include instances where LLMs output code into the chat interface, rather than in new files. In these instances, small instructions were additionally included to instruct models to generate the desired files, in an effort to make a more fair comparison between models that were and weren't finetuned for programming. 

As noted previously, the dataset of all websites generated for this study are included in our online GitHub repository\footnote{\url{https://github.com/TailiaReganMalloy/assessingSpearPhishing}} which additionally includes the internal reasoning output from LLM agents, and all analysis code used to generate the figures and statistics shown in the following sections. 

One of our main interests was how low or no experience users may employ these LLM coding agents in their attempt at spear-phishing. To that end we included in the prompts for models an image of the website that was to be copied. To ensure that models were willing to generate a copy of a website based on an image of the real site, we included in our prompt fake information contextualizing the request as being made by an instructor of an undergraduate course in cybersecurity. This contextualization of a positive outcome and educating students about cybersecurity principles was successful for all models in at least producing code, with no model refusing to generate a response or replying that it cannot follow our request. However, it is possible that more advanced systems were aware of the potential for misuse of the code that was generated and intentionally wrote poor quality code or a website that was visually very dissimilar. This possibility is explored in more depth in our section analyzing the internal reasoning produced by the models during the code generation requests. 

\newpage
\section{Results}
\setlength\LTleft{0pt}
\setlength\LTright{0pt}
\begin{longtable}{@{}p{0pt}@{}|@{\extracolsep{\fill}}c|c|c|c|c|c|c|@{}}
\hline
\ & Name & Context & Cost & Time & Tokens & Success & Similarity  \\
\hline
\endfirsthead
\hline
\ & Name & Context & Cost & Time & Tokens & Success & Similarity  \\
\hline
\endhead
\multirow{5}{*}{\makebox[0pt][r]{\rotatebox[origin=c]{90}{\textbf{Claude}}\hspace{0.8em}}} & Haiku-4.5 & 0.12M & \$0 & 0.06 h & 0.94M & 80\% & 17\%  \\
    \hline
 & Opus-4 & 0.11M & \$0 & 0.12 h & 0.37M & \textbf{\textit{100\%}} & 18\%  \\
    \hline
 & Opus-4.5 & 0.041M & \$0 & 0.00 h & 0.00052M & 0\% & N/A  \\
    \hline
 & Sonnet-4 & 0.12M & \$0 & 0.08 h & 1.2M & \textbf{\textit{100\%}} & 20\%  \\
    \hline
 & Sonnet-4.5 & 0.11M & \$0 & 0.05 h & 0.53M & 60\% & 23\%  \\
    \hline
    \hline
\multirow{5}{*}{\makebox[0pt][r]{\rotatebox[origin=c]{90}{\textbf{Gemini}}\hspace{0.8em}}} & 2.5-Pro & 0.092M & \$0 & 0.02 h & 0.19M & 60\% & 25\%  \\
    \hline
 & 3-Flash & 0.11M & \$0 & 0.02 h & 0.26M & \textbf{\textit{100\%}} & 21\%  \\
    \hline
 & 2.5-Flash & 0.82M & \$0 & 0.02 h & 0.43M & \textbf{\textit{100\%}} & 21\%  \\
    \hline
 & 3-Pro & 0.098M & \$0 & 0.05 h & 0.37M & \textbf{\textit{100\%}} & 21\%  \\
    \hline
 & 2.5-Light & 1M & \$0.45 & 0.03 h & 0.5M & \textbf{\textit{100\%}} & 18\%  \\
    \hline
    \hline
\multirow{5}{*}{\makebox[0pt][r]{\rotatebox[origin=c]{90}{\textbf{GPT}}\hspace{0.8em}}} & 4o & 0.063M & \$0 & 0.02 h & 0.23M & 40\% & 27\%  \\
    \hline
 & 5 & 0.11M & \$0 & 0.12 h & 0.84M & 80\% & 96\%  \\
    \hline
 & 5-Codex & 0.12M & \$0 & 0.17 h & 0.7M & 40\% & \textbf{100\%}  \\
    \hline
 & 5.1-Codex & 0.11M & \$0 & 0.22 h & 1.4M & \textbf{\textit{100\%}} & 54\%  \\
    \hline
 & 5.2-Codex & 0.23M & \$0 & 0.18 h & 0.53M & \textbf{\textit{100\%}} & 92\%  \\
    \hline
    \hline
\multirow{6}{*}{\makebox[0pt][r]{\rotatebox[origin=c]{90}{\textbf{Grok}}\hspace{0.8em}}} & 3 & 0.11M & \$0.84 & 0.06 h & 0.47M & 60\% & 20\%  \\
    \hline
 & 4 & 0.24M & \$0.97 & 0.09 h & 0.33M & \textit{80\%} & 12\%  \\
    \hline
 & 4-Fast & 2M & \$0.031 & 0.02 h & 0.19M & \textit{80\%} & 0\%  \\
    \hline
 & 4.1-Fast & 2M & \$0.024 & 0.04 h & 0.1M & \textit{80\%} & 25\%  \\
    \hline
 & Code-Fast-1 & 0.12M & \$0.2 & 0.23 h & 0.58M & 60\% & 14\%  \\
    \hline
    \hline
\multirow{5}{*}{\makebox[0pt][r]{\rotatebox[origin=c]{90}{\textbf{Llama}}\hspace{0.8em}}} & 3.1-70B & 0.12M & \$0.18 & 0.02 h & 0.25M & \textit{0\%} & N/A  \\
    \hline
 & 3.1-8B & 0.12M & \$0.2 & 0.25 h & 0.35M & \textit{0\%} & N/A  \\
    \hline
 & 3.3-7B & 0.12M & \$0.11 & 0.05 h & 0.14M & \textit{0\%} & N/A  \\
    \hline
 & 4-Maverick & 0.12M & \$0.19 & 0.05 h & 0.22M & \textit{0\%} & N/A  \\
    \hline
 & 4-Scout & 0.12M & \$0.12 & 0.01 h & 0.16M & \textit{0\%} & N/A  \\
    \hline
    \hline
\multirow{5}{*}{\makebox[0pt][r]{\rotatebox[origin=c]{90}{\textbf{Mistral}}\hspace{0.8em}}} & Large-3 & 0.22M & \$0.8 & 0.10 h & 0.89M & \textit{20\%} & 31\%  \\
    \hline
 & Medium-3 & 0.12M & \$0.03 & 0.00 h & 0.024M & 0\% & N/A  \\
    \hline
 & Medium-3.1 & 0.12M & \$0.57 & 0.05 h & 0.57M & 0\% & N/A  \\
    \hline
 & Small-3.2 & 0.12M & \$0.27 & 0.03 h & 0.41M & 0\% & N/A  \\
    \hline
    \hline
\multirow{3}{*}{\makebox[0pt][r]{\rotatebox[origin=c]{90}{\textbf{Nova}}\hspace{0.8em}}} & Lite-1 & 0.27M & \$0.32 & 0.05 h & 1M & \textit{0\%} & N/A  \\
    \hline
 & Micro-1 & 0.11M & \$0.076 & 0.01 h & 0.093M & \textit{0\%} & N/A  \\
    \hline
 & Pro-1 & 0.28M & \$0.31 & 0.03 h & 0.2M & \textit{0\%} & N/A  \\
    \hline
    \hline
\multirow{5}{*}{\makebox[0pt][r]{\rotatebox[origin=c]{90}{\textbf{Qwen}}\hspace{0.8em}}} & Coder-Flash & 0.11M & \$0.78 & 0.08 h & 0.75M & 0\% & N/A  \\
    \hline
 & Coder-Plus & 0.11M & \$1.7 & 0.09 h & 0.85M & \textbf{\textit{100\%}} & 31\%  \\
    \hline
 & Coder-3 & 0.25M & \$0.6 & 0.11 h & 1.2M & 0\% & N/A  \\
    \hline
 & Coder-30B & 0.14M & \$0.27 & 0.05 h & 0.39M & 0\% & N/A  \\
    \hline
 & 30B-Thinking & 0.17M & \$0.25 & 0.09 h & 0.36M & 0\% & N/A  \\
    \hline
	\caption*{}
	\label{tab:models}
\end{longtable}
\clearpage
\clearpage

\begin{figure}[ht]
  \centering
  \includegraphics[width=0.9\textwidth]{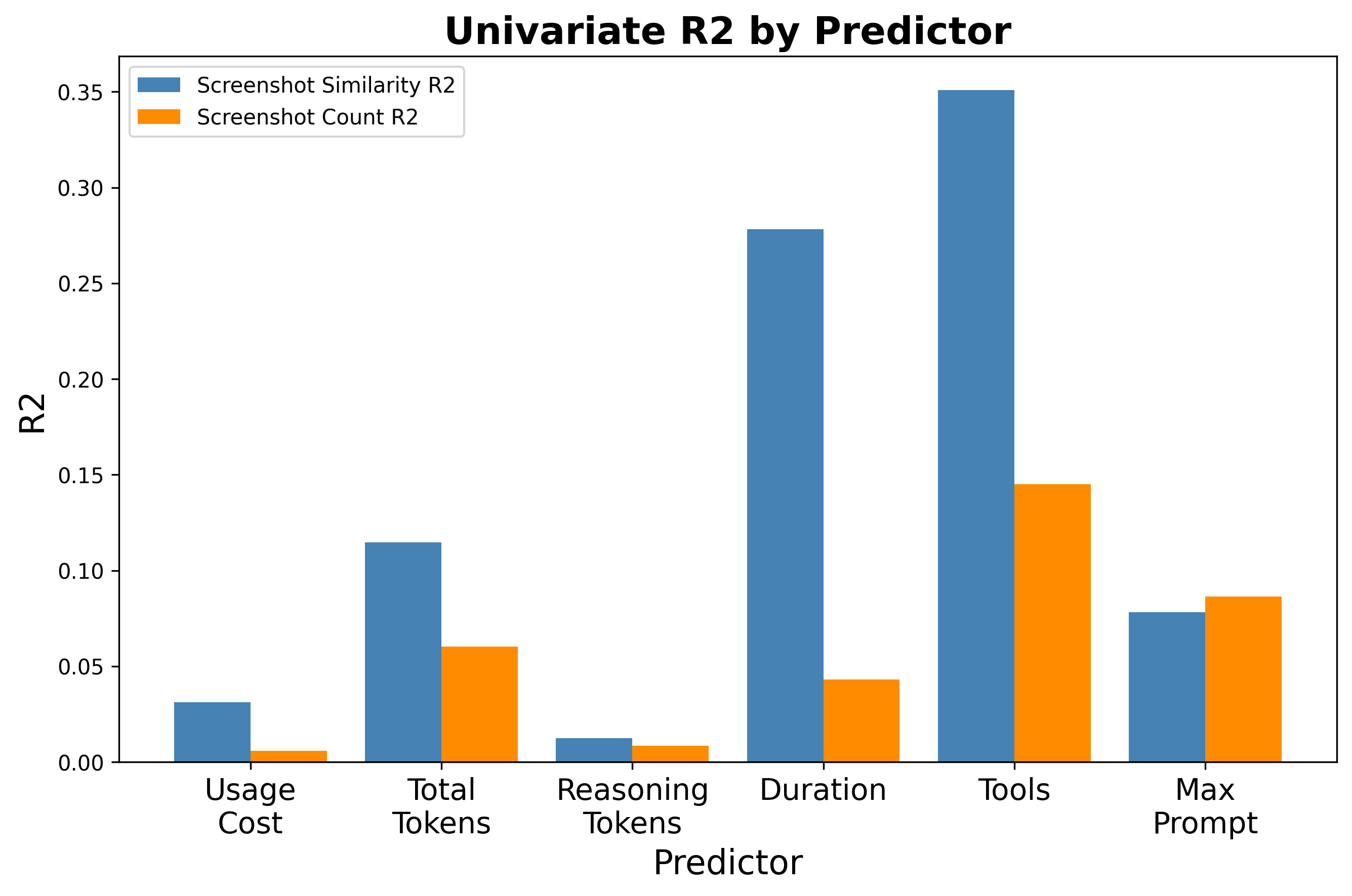}
  \caption{A comparison of the univariate R2 prediction of the screenshot similarity and the screenshot count by each of the six metrics under comparison.  } 
  \label{fig:RegressionComparison}
\end{figure}
\subsection{Model Performance Regression}
Figure \ref{fig:RegressionComparison} shows the results from 12 different univariate regressions by each of the 6 model metrics that we compare in Table \ref{tab:models}. Here we treat the number of screenshots that are generated by running the code bases and navigating to the local host as the success level of the models. This value ranges from 0 to 100 percent in increments of 20 percent because there are 5 possible screenshots that are made. The regression between each model feature and the number of screenshots that are generated is shown in orange. The next metric that we performed a regression on was the screenshot similarity. To make this similarity a percentage, the highest percent similarity of any of the websites that were generated was set to 100 percent and the lowest was set to 0 percent. This allows for a relative comparison across the different models of how successful they were in generating the target website. Even though the similarity metric is displayed as N/A in the above table, these values are not excluded from either predictor analysis and are set to 0. 

\subsection{Model Metrics}
The models used for evaluation are primarily grouped into the companies that generated these models, on the left hand side of Table \ref{tab:models} is shown the LLM families that are created by the different companies. The first column is the input prompt token size which is calculated by the average of the max tokens listed across all requests made during the generation of code. The next column is the cost of the requests made to the models in USD. Because the first three model families were built in to the MS Visual Studio Code co-piloting extension and were accessed using a github account with an educational license, the costs were not included in the json logs for the copilot chat debugging and cannot be estimated for individual models. The third metric of LLMs is the duration of the requests made during the internal reasoning. This is a good estimate of the overall time that it takes to generate all code, since the coding environments were set to automatically approve all commands from LLMs, meaning no input from humans was required during code generation. The fourth metric for comparison is the count of the total number of tools used by the model in code generation. 

During the prompting of models for code, there are several commands that are made available to them through tools. These include things like file creation and sending commands to the terminal. This metric is used to compare the relative ability of the models to perform as LLM Agents, rather than merely tools that can generate code but require intervention from human programmers. The final metric the the total number of tokens used by the model during all requests, this is closely related to the duration and usage but not exactly the same and is instead a metric of the complexity of internal reasoning. 

\subsection{Model Evaluation}
The evaluation of models is done using two values. The first, shown in the second to last column in the Table is the number of successful websites that are generated. Here, this is calculated by navigating into the code base that is generated and running a node command to start the local website. If the code runs with no errors and shows a webpage on the local host, it is counted as successful. Looking at this evaluation, we can see that the majority of success comes from the three built in models which are integrated into the copilot programming environment. Comparatively, only the Grok line of models reaches a similar level of success in runnable code compared to the other 4 model lines. However, many of the code bases generated by other model lines are close to being able to be run, and only require small changes, but this was avoided to better simulate a no-code task.  

Our of the models that generated runnable website code-bases, we next took screenshots with a  standardized aspect ratio and size and placed these images in the main folder of each code base. These screenshots were used with a simple similarity metric that compared the pixel level similarity of websites that were generated by LLM agents to the original source image that was used as a part of the prompt to make the websites. This was done to make the comparison quick and intuitive, since more complex comparisons like checking to see if certain website elements were or weren't present would require more work from humans. To fully automate the process we used this simple metric to compare the similarity of the websites to the original website. 

\newpage
\subsubsection{Model Max Prompt Size}
\begin{figure}[ht]
  \centering
  \includegraphics[width=0.9\textwidth]{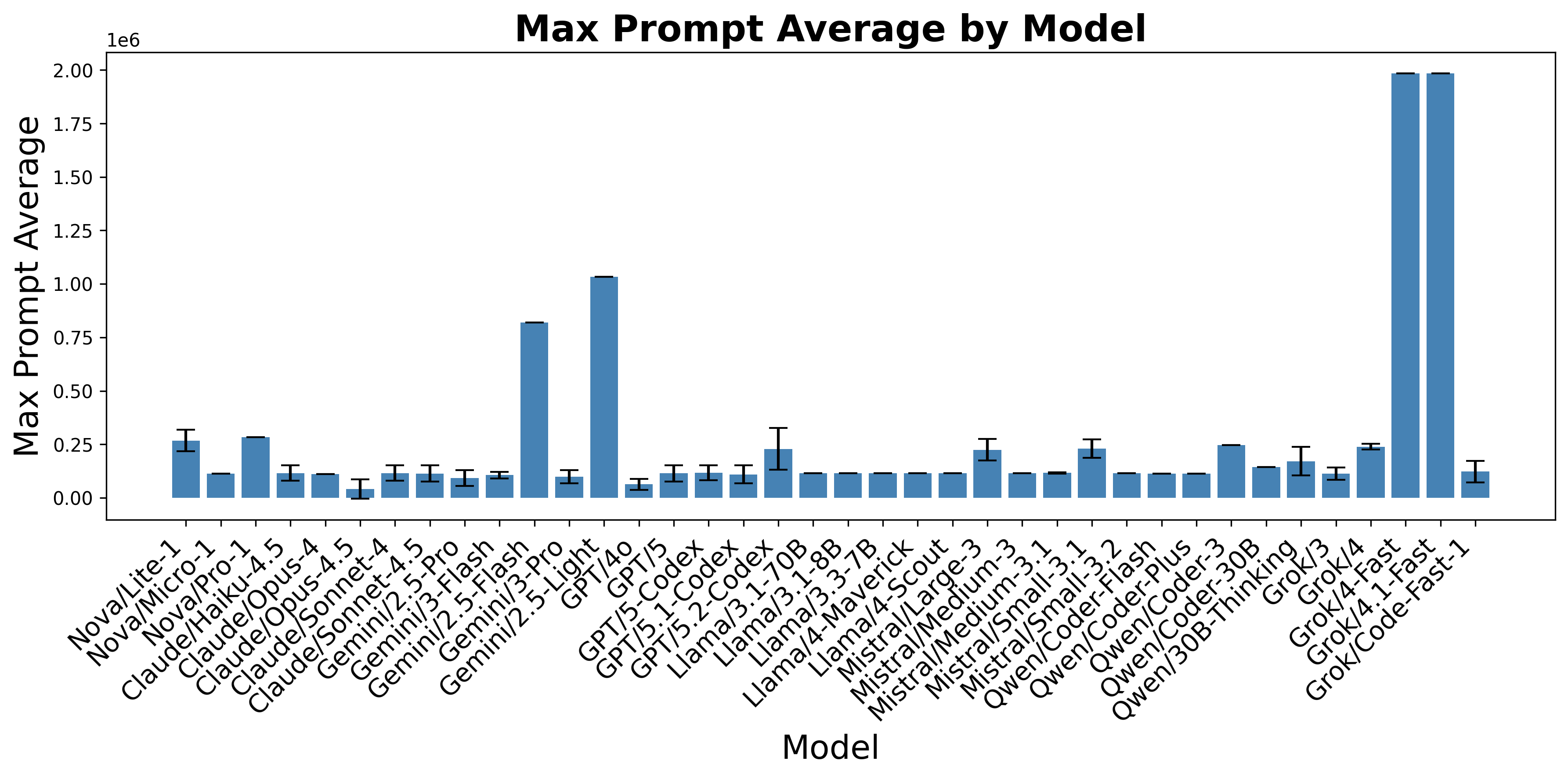}
  \caption{The max prompt size for each model averaged over all requests made during chain-of-thought reasoning while generating code. Brackets represent standard deviation of max prompt size.} 
  \label{fig:MaxPromptSize}
\end{figure}
The max prompt size or 'context' corresponds to the first column of Table \ref{tab:models}. In Figure \ref{fig:MaxPromptSize} we plot a bar chart to visually compare the different models under comparison in terms of the average max prompt size of the requests made during the internal prompting of the model. These internal promptings occasionally include prompts to the models with different max prompt sizes due to specific settings of the internal chain-of-thought reasoning processes of these models. Here, we can see the variation visually by comparing the error bars on the bar plots, which signify the standard deviation of the different max prompt averages across the five different prompts used to generate datasets. Overall, there is little variation across models in terms of the max prompt average, as well as within models in terms of their standard deviation. the four models that stand out are the Grok 4-Fast and 4.1 Fast as well as the Gemini 2.5 Flash and 2.5 Light. Interestingly, none of these are the models with the highest possible maximum prompt, indicating that the internal chain-of-thought reasoning sets different limits then the theoretical limits for the models. Figure \ref{fig:RegressionComparison} shows a comparison of the strength of the regression $r^2$ of a uni-variate regression on the size of the average max prompt in predicting the screenshot similarity and the screenshot success. This shows a moderate correlation compared to the other model metrics, with reasonable correlation for both the number of code bases that produced screenshots and the similarity of the generated websites. While models with more parameters can typically have larger possible maximum prompt size, this parameter is typically set lower than the actual maximum theoretical prompt of the model, instead limiting the prompt artificially for cost and time savings. 

\newpage
\subsubsection{Model Usage Cost}
\begin{figure}[ht]
  \centering
  \includegraphics[width=0.9\textwidth]{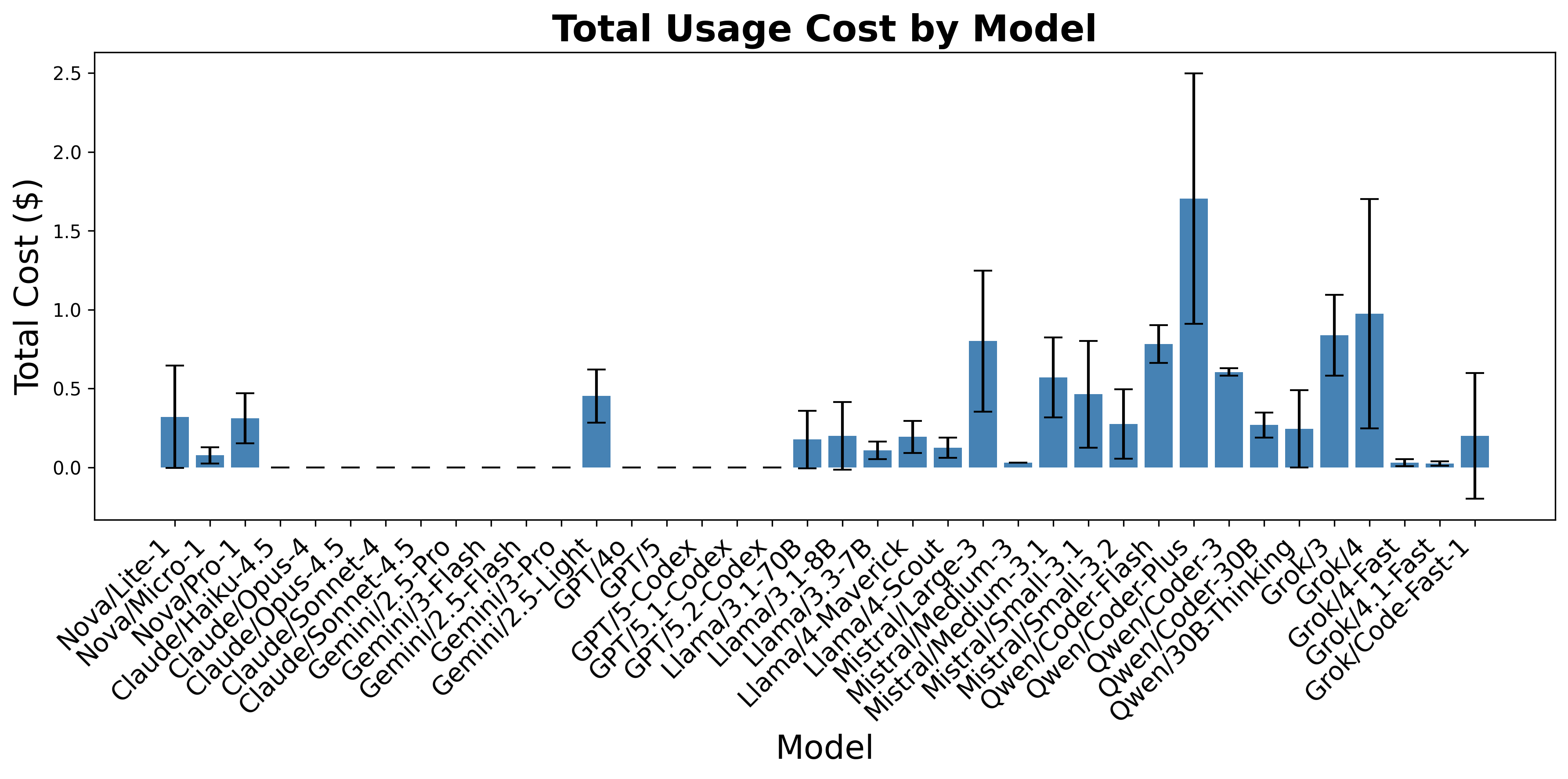}
  \caption{The total usage cost for each model totaled over all requests made during chain-of-thought reasoning while generating code. Brackets represent standard deviation of usage cost.} 
  \label{fig:ModelUsageCost}
\end{figure}
The second column of Table \ref{tab:models} compares the cost of the different models. In Figure \ref{fig:ModelUsageCost} we plot the total costs across models that reported their cost in chat logs.  As mentioned previously, the first three LLM families of GPT, Gemini, and Claude were all accessed using the MS Visual Studio Code extension Github Copilot, which relied on an educational account credit, which does not report individual prompt request costs within the chat logs. All other models were added onto this extension using OpenRouter, which does include the costs in their logs. This makes it difficult to compare using these results the abilities of the different models in terms of cost, since the majority of the non-builtin models did not generate runnable code.Additionally, an objective comparison of cost across all models is generally difficult due to differences in prices when accessing these models through the company's own API, or through third parities like the CoPilot extension and OpenRouter used here. However, the other model metrics under comparison are closely related to the cost, making it possible to draw some conclusions. Future research in LLM agent generated code that is interested in more directly comparing the impact of cost on code quality should use alternative methods of accessing models. 

We can see that there are many instances of models with high relative costs that were not able to generate runnable code for websites. This is likely an issue caused by the inability of these models to effectively interface with the programming environment, indicating that they could benefit from alternative methods of prompting or fine-tuning to better serve as LLM agents. Figure \ref{fig:RegressionComparison} shows a comparison of the strength of the regression $r^2$ of a uni-variate regression predicting the screenshot similarity and the screenshot success. This demonstrates that there is very little correlation between the cost and either success or similarity. 

\newpage
\subsubsection{Model Prompting Duration}
\begin{figure}[ht]
  \centering
  \includegraphics[width=0.9\textwidth]{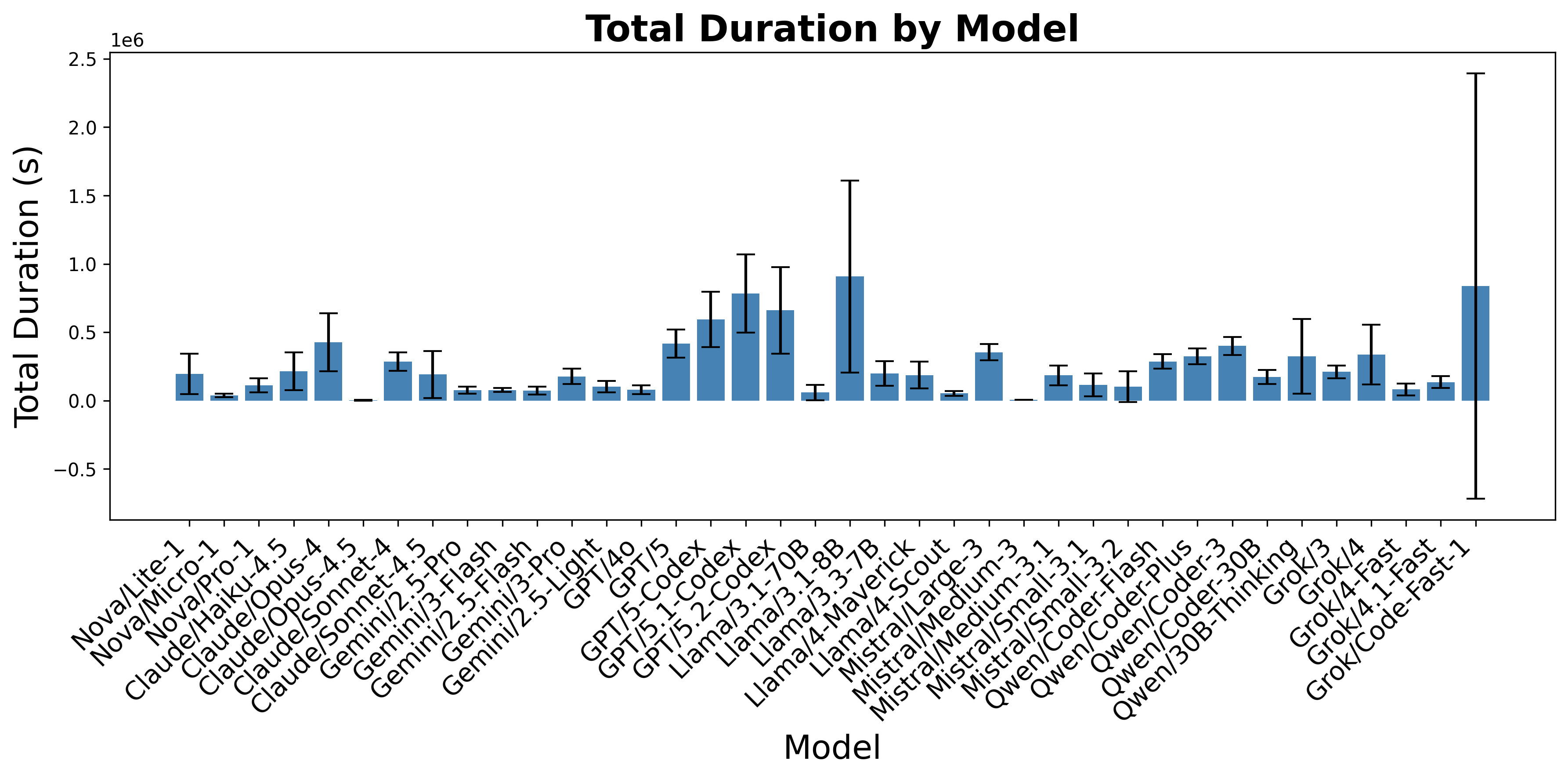}
  \caption{The chain-of-thought reasoning duration for each model totaled over all requests made while generating code. Brackets represent standard deviation of chain-of-thought reasoning duration.} 
  \label{fig:ModelPromptDuration}
\end{figure}
In Figure \ref{fig:ModelPromptDuration} we plot the total duration of chain-of-thought reasoning. One of the most variable model metrics is the total duration for the chain-of-thought prompting for the different models across the five different prompts. This is especially true of the models Grok Code Fast 1, Llama 3.1-8B and all of the GPT models. This is an interesting comparison as the prompts themselves did not include more or less instructions are were relatively similar in complexity. This demonstrates that initial small differences in prompts can produce significant differences in the length of time that it takes LLM Agent programmers to generate code. Another surprising aspect is found when comparing the model families that were better performing in terms of the number of websites that were generated without errors and the similarity of the generated websites. For instance, the GPT, Claude, and Gemini models were relatively successful, but only the GPT models demonstrated high chain-of-though reasoning time and variability. 

Comparing the regressions in Figure \ref{fig:RegressionComparison} shows an interesting difference between the correlation of model chain-of-thought reasoning duration and our two metrics. The relationship between duration and screenshot similarity is very high, the second highest of any of the different metrics. However, the relationship between the duration and the screenshot count is much lower, nearly the same as the lowest $r^2$ values for the screenshot count. This indicates that models that took longer to run were much more likely to generate websites that were similar to the original. Meanwhile, when comparing all models whether or not they produces runnable code, we can see that there is little correlation between the duration and the generation of code that runs without errors. Comparing the runtime of different models shows this is likely because of the high run times of models that did not end up producing code that ran without errors. 

\newpage
\subsubsection{Model Number of Tools Used}
\begin{figure}[ht]
  \centering
  \includegraphics[width=0.9\textwidth]{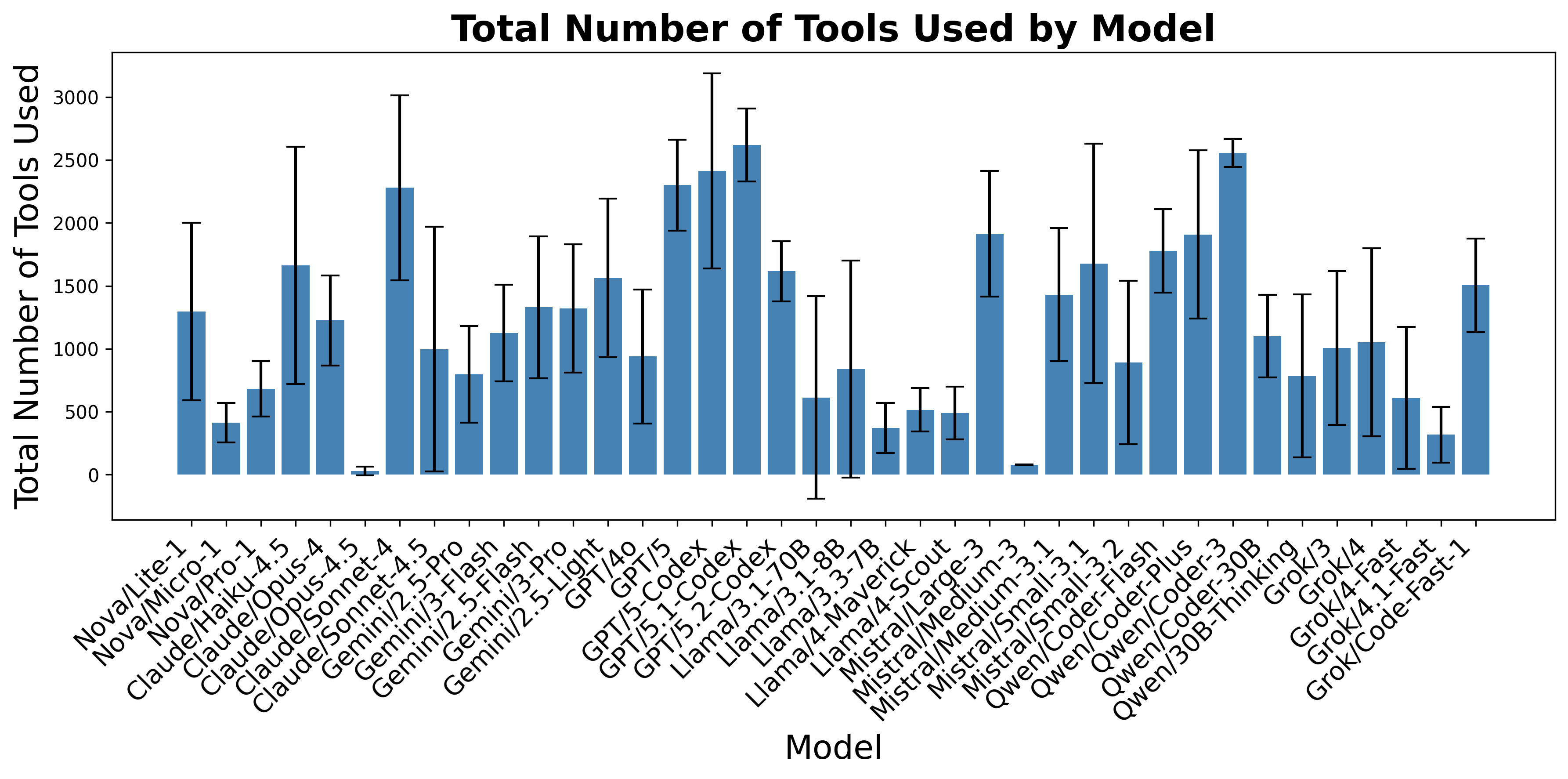}
  \caption{The number of tools used by each model totaled over all requests made during chain-of-thought reasoning while generating code. Brackets represent standard deviation of the number of tools used.} 
  \label{fig:ToolsUsed}
\end{figure}
Figure \ref{fig:ToolsUsed} shows the number of tool use requests that were made by different models during the chain of thought prompting from LLM agents during code generation. These tools include operations like creating files and running terminal commands. This is one of the main differences between LLMs as tools and LLM Agents that can perform actions through these tools. Comparing the number of tools used by the different models shows a wide variation within families of models, particularly Nova, Qwen and Claude. Additionally, there is a wide variance within specific model types, demonstrated by the wide error bars on most of the model tool use measurements. Comparing these tool uses does allow for a differentiation of models that refused to follow the prompt due to security and safety concerns, which were Claude Opus-4.5 and Mistral Medium-3. When these models were reasoning about the requests that were asked of them, they determined that the request was dangerous and did not get to the stage of making tool use requests to perform actions like creating documents or running terminal commands. 

When we compare the univariate R2 by predictor in Figure \ref{fig:RegressionComparison}, we can see that here is the highest R2 for tool use in predicting screenshot similarity and the number of websites that were successfully generated. One of the reasons for the high R2 for the number of screenshots that could be generated is the fact that some models refused to follow the prompt or, like the Llama family of models, produced an output that mostly describes how to generate the requested code without making as many tool requests. Additionally, models that were designed for code generation were much more successful in making tool command requests, even when comparing model families that did not frequently generate runnable code, such as Grok Code Fast 1. This indicates that encouraging more tool use could create better code bases, or additionally be used as a test to determine when LLM Agents are generating code copies. 

\newpage
\subsubsection{Model Total Tokens Used}
\begin{figure}[ht]
  \centering
  \includegraphics[width=0.9\textwidth]{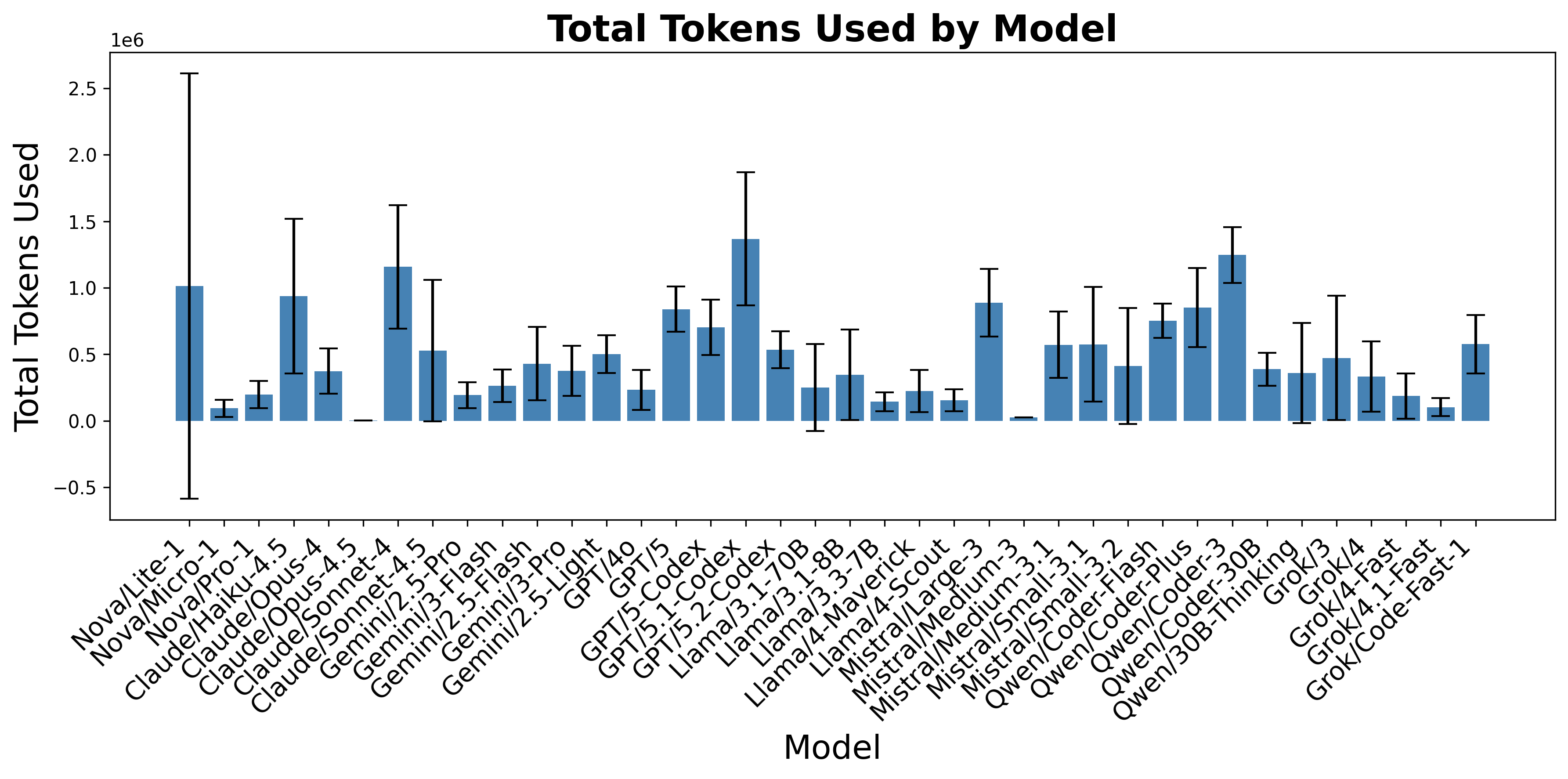}
  \caption{The number of tokens used by each model totaled over all requests made during chain-of-thought reasoning while generating code. Brackets represent standard deviation of the number of tokens used.} 
  \label{fig:totalTokens}
\end{figure}
The final metric that we compare in Figure \ref{fig:totalTokens} is based on the total number of tokens made in all of the requests that are sent during the chain of thought reasoning that LLM Agents used to generate the code bases. Comparing this graph to the previous one showing the number of tools used by models shows that, unsurprisingly, models that used more tools also had more total requests. However these is less variation in the total number of tokens used by models both within individual models and in model families. As with the number of tools used, we can see the two models that refused to perform the task, Claude Opus 4.5 and Mistral Medium 3. However we can also see that comparatively low total tokens are used by models that attempted to generate websites but were unsuccessful, such as the Llama family of models. Interestingly, there appears to be roughly the same level of total tokens in the Llama and Gemini families, even though they had very different performances in generating website code bases that would run. This could be due to the fact that Llama models continue self-prompting during reasoning when they are unsuccessful in producing correctly formatted commands to produce the tool use.

This effect of two groups of models with similar total tokens used but very different performance in terms of producing useful code is demonstrated when comparing the R2 values of the total tokens used by models and the number of screenshots and the screenshot similarity in Figure \ref{fig:RegressionComparison}.

\section{Discussion}
In this work, we describe a pipeline for evaluating LLM coding agents in their potential for misuse in spear-phishing attacks through the generation of phishing websites. The results from this assessment include a comparison of 5 different prompts with 40 LLMs from 8 companies. We measured the performance of these models in terms of the percentage of prompts that generated code that could be used to run a website, and how similar the generated websites were to the target input website. The goal of this research was to present a method for assessing how willing and able different LLM models are in generating spear-phishing websites. To do this, we designed a pipeline that could be used to generate several different code bases all attempting to copy the same target website. We applied this pipeline onto an evaluation of a real-world spear-phishing website generation task, by prompting models to generate a clone of a specific website. The majority of the analysis that we perform in this work focused around the different metrics for LLM models and how they impacted our two measures of coding capability. We also include a dataset of LLM generated phishing websites that can be useful for future researchers interested in LLM safety and trustworthiness, as well as LLM applications in cybersecurity and social engineering. 

Using six different metrics of LLMs we compared the correlation between these metrics and our two performance metrics. This correlation analysis showed that models that took more time and used more programming tools to generate websites were much more likely to generate similar looking websites. Meanwhile, few of these metrics were strongly correlated with the success of the models in generating runnable code. Two of the metrics that had surprisingly low correlations with performance was the max prompt size of the model, which is correlated to overall model size, and the usage cost. Part of the reason for this was that many of the better performing models are made available to programmers for free. Additional future research can better compare the correlation between cost and performance in cybersecurity coding contexts by altering the method of prompting LLMs.

An important issue in LLM research is that of reproducibility. To allow for better reproducibility of the results we analyze, we have included in our published dataset information about all of the models used to generate code. This includes information such as the specific model versions, the dates used to access them, as well as input prompts, methods of access, and parameters that are set by the Copilot programming extension like the model temperature. The stochastic nature of these models means that perfectly replicating our method of generating these datasets likely won't generate the same code. However, this issue of stochasticity exists in all studies of the output of LLMs, even when setting temperature parameters to zero the output of these models is never fully deterministic. Another complication is that many of the models we use are accessed through APIs that are not completely transparent in how the models are prompted and what pre-input prompts or alterations to prompts are done. This is particularly relevant in the analysis of which models refused to generate the spear-phishing website code, since it is likely that these systems are at least in part responsible for the LLM coding agents refusal to generate spear-phishing websites. Overall, it is important to think of these analyses as a snapshot of the behavior of LLM agents in time that, alongside continued research and analysis of LLM agent behavior, can build a picture of how LLM coding agents behave and how this can be exploited in cybersecurity contexts.

While our analysis provided interesting insight into the potential misuse of LLM coding agents in cybersecurity, there are several defensive applications that this work applies to. Firstly, we present a large set of different code bases that were generated by a wide variety of LLM coding agents, all centered around the same goal and set of prompts. Alongside these code bases, we also include in our dataset the internal reasoning performed by the LLM agents during code generation. We hope to apply these datasets in the future in two main areas for improved defense against LLM agent misuse in spear-phishing. Firstly, by building a tool such as a website extension that can compare the source code of websites against an existing database of LLM generated spear-phishing websites, to prevent users from accessing these potentially dangerous sites. Secondly, we hope to perform further analysis on the internal reasoning of the LLM agents when they are generating code, to design reasoning strategies that can better align with safety while also being useful educational tools.

\bibliography{springer}

\clearpage

\end{document}